\documentclass[12pt,preprint]{aastex}
\usepackage{emulateapj5}

\usepackage{psfig}
\usepackage{graphicx}
\usepackage{times}



\newcommand{\lsim}{\mbox{\hspace{.2em}\raisebox{.5ex}{$<$}\hspace{-.8em}\raisebox{-.5ex}{$\sim$}\hspace{.2em}}}
\newcommand{\gsim}{\mbox{\hspace{.2em}\raisebox{.5ex}{$>$}\hspace{-.8em}\raisebox{-.5ex}{$\sim$}\hspace{.2em}}}

\def\chandra    {{\em Chandra}\/}

\def\einstein   {{\em Einstein}\/}

\def\rosat      {{\em ROSAT}\/}

\def\hst        {{\em HST}\/}

\usepackage{mathptmx}

\begin{document}


\title{The survival and destruction of X-ray coronae of early-type galaxies in the rich cluster environments: a case study of Abell 1367}

\author{M.\ Sun,
  A.\ Vikhlinin, 
  W.\ Forman,
  C.\ Jones,
  S.\ S.\ Murray}
\affil{Harvard-Smithsonian Center for Astrophysics, 60 Garden St.,
Cambridge, MA 02138;\\ msun@cfa.harvard.edu}

\shorttitle{X-ray galaxy coronae in A1367}
\shortauthors{SUN ET AL.}

\begin{abstract}
A new \emph{Chandra} observation of the northwest (NW) region of the
galaxy cluster A1367 reveals four cool galaxy coronae (0.4 - 1.0 keV)
embedded in the hot intracluster medium (ICM) (5 - 6 keV). While the
large coronae of NGC~3842 and NGC~3837 appear symmetric and relaxed,
the galaxy coronae of the $\lsim$ L$^{\ast}$ galaxies (NGC~3841 and
CGCG~97090) are disturbed and being stripped. Massive galaxies, with
dense cooling cores, are better able to resist ram pressure stripping
and survive in rich environments than $\lsim$ L$^{\ast}$ galaxies
whose galactic coronae are much less dense.
The survival of these cool coronae implies that
thermal conduction from the hot surrounding ICM has to be suppressed by a
factor of at least 60, at the corona boundary. Within the galaxy coronae
of NGC~3842 and NGC~3837, stellar mass loss or heat conduction with the
Spitzer value may be sufficient to balance radiative cooling. Energy
deposition at the ends of collimated jets may heat the outer coronae, but
allow the survival of a small, dense gas core (e.g., NGC~3842 in A1367 and
NGC~4874 in Coma). The survived X-ray coronae become significantly smaller
and fainter with the increasing ambient pressure.
\end{abstract}


\section{Introduction}

Galaxy coronae with temperatures of $\sim$ 10$^{7}$ K are common in early-type
galaxies (Forman, Jones \& Tucker 1985). \einstein\ and \rosat\ observations
led to the detection of many such coronae in the field, galaxy groups and poor
clusters (e.g., Brown \& Bregman 1998; Beuing et al. 1999). However, prior to the launch of
\chandra, no 10$^{7}$ K galaxy corona were detected in hot (T $\sim$ 10$^{8}$
K) clusters, nor were they expected, since evaporation and ram-pressure
stripping by the hot and dense ICM should be very efficient. The first direct
evidence for the survival of galaxy coronae in hot clusters came from the
\chandra\ observations of the Coma cluster. Vikhlinin et al. (2001, V01) found
small but extended ($\sim$ 2 kpc in radius for h$_{0}$=0.7) X-ray coronae
(T $\sim$ 1-2 keV) in the two dominant Coma galaxies, NGC 4874 and NGC 4889,
while the surrounding ICM has a temperature of 8 - 9 keV. The existence of
these cool coronae implies that heat conduction must be suppressed by a factor
of 30 - 100 at the ISM-ICM boundary. Yamasaki, Ohashi \& Furusho (2002) also
reported two small 0.7 - 1.1 keV galaxy coronae ($\sim$ 2 - 3 kpc in radius
for h$_{0}$=0.7) in the center of A1060 cluster, although the ICM is not hot
(3 keV). These galaxy coronae in rich environments (high ambient pressure) are
generally small (2 - 3 kpc, or 4 - 7$''$ in radius at the distance of the Coma
cluster) and require the superior \chandra\ angular resolution to resolve them.

These embedded galaxy coronae are perfect targets to study the effects of
rich environments on galaxy coronae. The survival of galaxy coronae in rich
environments is difficult, mostly due to gas stripping and evaporation
by the surrounding hot dense ICM. These processes may be very efficient
and may fully destroy the coronae. Thus, the properties of galaxy coronae
in hot clusters and the frequency of their existence, in combination with
the properties of galaxies without coronae, provide valuable information
on gas stripping processes, stellar mass loss rates in early-type galaxies
and the suppression of heat conduction. In principle, the properties of
coronae also shed light on the evolution of their early-type host galaxies,
as the evolution processes (e.g., mergers) and the environment can impact
the evolution of the coronae. It is also interesting to explore whether
galaxy coronae in hot clusters can serve as seeds of larger cool cores
(e.g., Motl et al. 2004).

A 40 ks ACIS-S observation of A1367 revealed two candidate X-ray coronae
associated with the elliptical galaxies NGC 3842 and NGC 3837 (Sun \& Murray
2002b; S02b hereafter). Both galaxies lie in a merging subcluster 18$'$ (or
450 kpc) NW of the X-ray peak of A1367 (Fig. 1). This region is
significantly hotter ($\sim$ 5 keV) than the rest of the cluster ($\sim$ 3
keV, Donnelly et al. 1998; Sun \& Murray 2002a; S02a hereafter), which can
be explained by shock heating from an ongoing merger of the NW subcluster
with the primary cluster. NGC~3842 is the brightest galaxy at the center of this
subcluster, although it has no extended optical envelope. NGC~3837 is the 
second brightest early-type galaxy in the subcluster. With a new \chandra\
observation centered on the subcluster, we resolve these two sources into
extended 10$''$ galaxy coronae with temperatures of 0.6 - 1.0 keV. Two smaller
and amorphous galaxy coronae are also detected to be associated with smaller
galaxies NGC~3841 and CGCG~97090. We discuss the physical effects that
influence these coronae.

Throughout this paper we assume H$_{0}$ = 70 km s$^{-1}$ Mpc$^{-1}$,
$\Omega$$_{\rm M}$=0.3, and $\Omega_{\rm \Lambda}$=0.7. We use the
redshift of the subcluster central galaxy NGC~3842, z=0.02107, to
calculate the luminosity distance, 91.9 Mpc. 1$''$ corresponds to
0.428 kpc. We use the Galactic absorption of 2.2$\times$10$^{20}$
cm$^{-2}$. The solar photospheric abundance table by Anders \& Grevesse
(1989) is used in the spectral fits. Uncertainties quoted are 1 $\sigma$.

\section{\chandra\ observations}

\subsection{\chandra\ data reduction}

A 48 ksec \chandra\ observation was performed with the Advanced CCD Imaging
Spectrometer (ACIS). The optical axis is on the CCD S3 and the
data were telemetered in Very Faint mode. Standard data analysis
is performed which includes the correction for the slow
gain change\footnote{http://cxc.harvard.edu/contrib/alexey/tgain/tgain.html}.
The science results in this paper are mainly from the CCD S3. We investigated
the light curve from the CCD S1 which is far from the cluster center and
where the background is dominant. Excluding time intervals with significant
background flares, resulted in a total exposure of 43.2 ksec. For the studies
of small galaxy coronae, the backgrounds are from their immediate surroundings.
For the analysis of the cluster diffuse emission, we used the period
D background file, ``aciss\_D\_7\_bg\_evt\_271103.fits''
\footnote{http://cxc.harvard.edu/contrib/maxim/bg/index.html}.
As pointed out by Markevitch (2002), weak background flares
in the S3 CCD can significantly bias the temperature measurement.
Thus, for the analysis of cluster diffuse emission, we used the same strict
criterion of excluding background flares as was used to produce the background
data files. For this analysis, the effective exposure is 36.8 ksec.
The particle background levels (measured in PHA channels 2500-3000 ADU)
were 4.8\% lower than that of the period D background data. Thus, we
decreased the background normalization by 4.8\%.

We made the correction for the ACIS low energy quantum efficiency (QE)
degradation, which increases with time. The calibration files used
correspond to CALDB 2.28 from the \chandra\ X-ray Center (CXC),
released on August 11, 2004. In the spectral analysis, a low energy
cut of 0.5 keV is used.

\subsection{Global properties of the A1367 galaxy coronae}

The \chandra\ raw image of the region around NGC~3842 is shown in Fig. 1,
where all four galaxy coronae are shown. The two bright
coronae associated with NGC~3842 and NGC~3837 are resolved. Moreover,
two smaller and fainter extended sources, associated with early-type galaxies
NGC~3841 and CGCG~97090, are detected. The X-ray colors of these sources
indicate that they are much softer than expected for low-mass X-ray
binaries (LMXB) and are consistent with thermal coronae. Basic properties
of these galaxies are given in Table 1 and their X-ray contours are plotted
in Fig. 2. The X-ray emission from NGC~3842 is symmetric around the nucleus
and the X-ray peak is within 0.1$''$ of the optical peak determined from \hst\
WFPC2 observations. The NGC~3837 corona is symmetric within the central
8$''$. The peak of the NGC~3837 corona is $\sim$ 1.1 $''$ south of the
optical peak, which is consistent with the positional uncertainties of the
DSS and \chandra. Although only 30\% fainter than NGC~3837 optically,
NGC~3841 is 9 times fainter in X-rays. This may be related to its
considerably smaller velocity dispersion than NGC~3837 (Table 1), which
generally implies a significantly smaller total mass for NGC~3841. 
The peak of the NGC~3841 corona is $\sim 1.9''$ north of
its optical peak, which suggests that its gas core has been displaced by ram
pressure. The X-ray peak of the CGCG~97090 corona is consistent with the
\hst\ center within 0.1$''$, but its X-ray emission is extended to the NW.
Interestingly, the directions of motion of the off-center galaxies,
derived from the displacement and extension of the X-ray coronae, are all towards
or near the direction of the central galaxy NGC~3842.

The diffuse cluster emission around these coronae can now be better understood,
without the contamination of bright point sources and coronae suffered for
the \rosat\ PSPC data (Fig. 1). The X-ray emission from coronae, point
sources and UGC~6697 (the peculiar galaxy to the west of NGC~3842) is masked
and the smoothed image of the cluster diffuse emission is shown in Fig. 3. The
cluster emission is elongated to the southeast (SE) and there is only a weak
X-ray concentration around NGC~3842 (4-5 $\sigma$ significance). No other
significant structure in the cluster diffuse emission is found.

The 0.5 - 2 keV surface brightness profiles around NGC~3842 and NGC~3837
are shown in Fig. 4. The X-ray emission of the NGC~3842 and NGC~3837 coronae
is composed of two components: a bright
steep central core within 9.6$''$ for NGC~3842 and 6.8$''$ for NGC~3837,
and a weak flat component from 9.6$'' - 24.0''$ for NGC~3842 (5.8
$\sigma$) and 6.8$'' - 17.2''$ for NGC~3837 (3.9 $\sigma$) (``plateau'').
The light from LMXB may be responsible for the ``plateau'' regions, which
is discussed in $\S$2.3.

Integrated spectra of the coronae were extracted within 24.0$''$ for NGC~3842
and 17.2$''$ for NGC~3837. Background spectra were collected in the 24.0$''$ -
51$''$ and 17.2$''$ - 40$''$ annuli around NGC~3842 and NGC~3837 respectively.
The surface brightness around NGC~3842 (24.0$''$ - 70$''$) shows
a small condensation (the enlarged region of Fig. 4). We fit the profile with a
power law and rescale the normalization of the background spectrum using the
extrapolation of the power law fit. The ICM background around NGC~3837 is
however very flat. The background-subtracted spectra of the two coronae
are shown in Fig. 5 with the best-fit MEKAL model. The blend of iron L lines
is significant. The spectra of the coronae and the surrounding ICM emission
are fitted with MEKAL model. As seen from the fits
in Table 2, both coronae are indeed much cooler (0.6 - 0.9 keV) than their
surroundings (5 - 6 keV), which is consistent with our previous results (S02b).
The best-fit average abundance is kind of low (0.3 solar). The 90\% higher
limit on the average abundance is 0.53 and 0.63 for NGC~3842 and NGC~3837
respectively. In view of the possible Si line in the spectrum of NGC~3842,
we also fit its spectrum with a VMEKAL model. The derived Si abundance is
0.88$^{+0.45}_{-0.33}$ solar. For iron L blend, spectral components with
different temperatures may smooth the blend and make the accurate determination
of iron abundance difficult, especially when the data statistics are not good.
This issue may only be resolved with the high spectral resolution data.

For massive elliptical galaxies, we may expect significant X-ray emission from
the central source and LMXB. There is no central X-ray point source detected
in either NGC~3842 or NGC~3837 and the upper limit of the 0.5 - 10 keV
luminosity is 4 (NGC~3842) and 5 $\times 10^{39}$ ergs s$^{-1}$ (NGC~3837) respectively.
For their optical luminosities, we expect 0.3 - 10 keV total luminosities
of 2.2 and 1.0 $\times 10^{40}$ ergs s$^{-1}$ for the LMXB components
within the integrated spectra of NGC~3842 and NGC~3837 (note the background
spectra also contain some LMXB flux), based on the L$_{\rm LMXB}$ - L$_{\rm B}$
scaling relation from Sarazin, Irwin \& Bregman (2001). The current data
show hard component on the significance level of 2.0 $\sigma$ and 3.0 $\sigma$
in the integrated spectrum of NGC~3842 and NGC~3837 respectively (10 - 20 \%
more significant in the outskirts of coronae). The derived temperatures of the
galaxy coronae change little when this hard component is included (Table 2). If we
assume a temperature of 8 keV (Sarazin et al. 2001), the derived 0.3 - 10 keV
luminosities of the hard component are 2.7$^{+1.5}_{-1.3} \times 10^{40}$
ergs s$^{-1}$ (NGC~3842) and 2.0$\pm$ 0.6 $\times 10^{40}$ ergs s$^{-1}$
(NGC~3837). As the L$_{\rm LMXB}$ - L$_{\rm B}$ scaling relation has a
dispersion on the level of 60\% (Kim \& Fabbiano 2004) and the total luminosity
of LMXB is mainly determined by those brightest sources, the above
values are consistent with the expectation.

The spectra of the two fainter coronae (NGC~3841 and CGCG~97090) were also
extracted though the statistics are poor. If they are fitted with a single
thermal model (MEKAL), the derived temperatures are low but with large
uncertainties (Table 1). The abundances are unconstrained. In addition,
there may be significant contribution from LMXB and nuclear sources.
The LMXB are expected to contribute $\sim$ 1/4 and 1/8 (with 60\% uncertainty)
of the 0.5 - 2 keV emission within the corona of NGC~3841 and CGCG~97090
respectively. The contribution of the nuclear source to the X-ray emission of
NGC~3841 is small since its X-ray peak is 2$''$ north of the nucleus. However,
the nuclear source may contribute $\sim$ 1/4 of X-ray light from CGCG~97090.
Thus, up to 40\% of the X-ray emission in NGC~3841 and CGCG~97090 may not
come from thermal gas. Assuming constant density for the coronae of NGC~3841
and CGCG~97090 and an average abundance of 0.3 solar, the average electron
density is $\sim$ 0.02 cm$^{-3}$ and the total gas mass is 10$^{6} - 10^{7}$
M$_{\odot}$.

There are three other member galaxies (NGC~3840 - Sa, NGC~3844 - S0/a,
and NGC~3845 - S0) detected as faint X-ray point-like sources at the position
of nuclei, all in the S2 CCD. Their hardness ratios (1.4 - 5.0 keV /
0.7 - 1.4 keV) suggest much harder emission than the cool coronae discussed
above. They are most likely low luminosity AGN with 0.5 - 10 keV X-ray
luminosities of $\sim 10^{40}$ ergs s$^{-1}$.

\subsection{Radial properties of the NGC~3842 and NGC~3837 coronae}

As shown in Fig. 4, there are two components in the surface brightness
profiles of NGC~3842 and NGC~3837 coronae. The flat one may be related
to the LMXB component of the galaxy. Assuming the LMXB light follows
the optical light, the expected LMXB light profile of NGC~3842 can be
derived based on \hst\ WFPC2 observations (Fig. 4). The normalization
is determined to match the expected total luminosity ($\S$2.2). There is
no \hst\ data for NGC~3837. We extracted its B-band light profile from a
CFHT image obtained from GOLD Mine\footnote{http://goldmine.mib.infn.it}.
Although the ``plateau'' region in NGC~3837 may also arise from LMXB, its
X-ray emission is more extended to the south (Fig. 2), which is confirmed
by the surface brightness profiles in the north and south for the 6.8$''$ -
17.2$''$ region (Fig. 4). Thus, the emission from the stripped envelope
contributes to most, if not all, emission in the ``plateau'' region of NGC~3837.
For detailed studies of the inner profiles, we must account for the effect
of the \chandra\ PSF since it is not negligible compared to the source size,
especially for NGC~3837, where the PSF is elongated NW-SE.
The PSF model was obtained from ChaRT simulations and the source spectra
were taken as inputs to obtain the energy-dependent PSF. We then performed
SHERPA 2D fits to the exposure-corrected and background-subtracted images,
with the PSF correction at the position of the sources. Both the ICM and
estimated LMXB components were subtracted. The assumed corona model is
a 2D $\beta$ model with a brightness cut at r$_{\rm cut}$, which well
describes the inner component. The results are listed in Table 3.

We derived the temperature profiles of the NGC~3842 and NGC~3837 coronae.
The background spectra are from the surrounding ICM ($\S$2.2). In each
annulus, we fit the spectrum with a MEKAL model. An additional 8 keV thermal
bremsstrahlung component (ZBREMSS), representing the contribution from LMXB, 
is added if there is a clear hard X-ray excess in the spectrum. The
inclusion of this hard component decreases the best-fit temperatures by less
than 8\%. The results are presented in Table
2 and the temperature profiles are shown in Fig. 4. The temperatures of the
``plateau'' regions are poorly determined, but are still consistent with
the values expected from the LMXB component. The ICM temperatures around the coronae
are consistent with the average temperature of the region (5.3$\pm$0.2 keV from
the integrated cluster spectrum on the S3 CCD). This high temperature is consistent
with previous measurements (Donnelly et al. 1999; S02a).

From the best-fit $\beta$ models to the inner profiles of the coronae, the
electron density distribution and gas mass within r$_{\rm cut}$ can be
derived (Table 3). The emissivity of the gas is assumed to be the same
and the value derived from the fit to the global spectrum is used. The
central density is indeed high ($\sim$ 0.2 cm$^{-3}$) and the gas mass
is $\sim 10^{8}$ M$_{\odot}$. Current data cannot exclude the possibility
of the existence of the thermal gas in the ``plateau'' region of MGC~3842
and the stripped gas may be a significant component in the ``plateau''
region of MGC~3837. The gas mass in the ``plateau'' region can be comparable
to the total gas mass within r$_{\rm cut}$.

It is important to know the density of the ICM surrounding these coronae.
NGC~3842 is the central brightest galaxy of the NW subcluster and
should be located close to the bottom of the potential well, which
is supported by a surrounding arcmin-scale ICM enhancement (Fig. 3 and 4).
Assuming a constant density, we derive the ICM electron density
around NGC~3842 of $\sim 1.1 \times 10^{-3}$~cm$^{-3}$.
The uncertainties on the geometry and gas distribution give a range of
0.9 - 1.5 $\times 10^{-3}$~cm$^{-3}$.
The ICM density around NGC~3837 is more uncertain because of its unknown
location along the line of sight. The $\beta$-model fit to the NW subcluster
by Donnelly et al. (1998) gives an electron density of $\sim 6 \times 10^{-4}$
cm$^{-3}$ at the projected position of NGC~3837. We adopt an uncertainty
of 30\%.

\section{Discussion}

\subsection{The stripping of X-ray coronae in the cluster}

Within the NW subcluster the galaxy coronae face gas stripping
if the galaxies are moving significantly relative to the surrounding ICM.
The galaxy line-of-sight velocity dispersion derived from a sample of 26
galaxies within 300 kpc (or 12$'$) radius of NGC~3842 is $\sim$ 840 km/s,
which is similar to the velocity dispersion of the whole cluster
(822$^{+69}_{-55}$ km/s from Zabludoff, Huchra \& Geller 1990). However,
the central galaxy NGC~3842, shielded by the associated arcmin-scale ICM
enhancement (Fig. 3), may have small residual motion relative to the
surrounding ICM. The effects of stripping also depend on the internal
properties of the coronae. For the most massive early-type galaxies, both
observations and theories (e.g., David, Forman \& Jones 1991; Pellegrini \&
Ciotti 1998) indicate that a galactic cooling core has developed, which is
hard to be stripped because of the high gas density. The smaller galaxies
(L$_{\rm B} \lsim 10^{10}$ L$_{\odot}$)
have much less hot gas (e.g., L$_{\rm X} \propto$ L$_{\rm B}^{3.0-3.5}$,
Brown \& Bregman 1998; L$_{\rm X} \propto$ L$_{\rm B}^{\sim 2.7}$, O'Sullivan,
Ponman \& Collins 2003). Their X-ray coronae are in the wind or partial wind
stage (David et al. 1991; Pellegrini \& Ciotti 1998) so that the gas density
is low. These small galaxies are more vulnerable to ram pressure stripping.

The gas stripping can occur in the following ways: prompt ram pressure stripping,
continuous stripping by the Kelvin-Helmholtz (K-H) instability and viscosity
(e.g., Nulsen 1982; Takeda, Nulsen \& Fabian 1984; Toniazzo \& Schindler 2001).

The criteria for prompt ram pressure stripping of a galaxy corona (gas density
$\rho_{\rm ISM}$) moving with a velocity (v$_{\rm gal}$) in the ICM (gas density
$\rho_{\rm ICM}$) is given by:

\begin{equation}
\rho_{\rm ICM}{\rm v}_{\rm gal}^{2} \gsim \int \rho_{\rm ISM} {\bf \nabla \phi_{\rm gal}} \cdot d{\bf l}
\end{equation}

\noindent
where $\phi_{\rm gal}$ is the galactic potential and the integral is taken along
the direction of motion.

We first estimate the critical velocity to fully strip the coronae of NGC~3842,
as an example of luminous galaxies. We assume M$_{\ast}$ / L$_{\rm B}$ = 7
M$_{\odot}$ / L$_{\odot}$ and use the stellar mass potential (from the \hst\
data) to represent the total mass potential within the small corona. The gas
density profile of the NGC~3842 corona is known ($\S$2.3).
The derived critical velocity to fully strip the X-ray corona of NGC~3842 is:
1.8 $\times 10^{3}$ ($\frac{n_{\rm ICM}}{1.1\times10^{-3} cm^{-3}})^{-1/2}$
km s$^{-1}$, which is twice of the velocity dispersion of the cluster. When
stellar mass loss is included, complete stripping is even more difficult (e.g.,
Takeda et al. 1984). Although there is no \hst\ data for NGC~3837, assuming a simple
r$^{1/4}$ law, the critical velocity to fully strip the coronae of NGC~3837 is
comparable to that of NGC~3842. Thus, in the A1367 environment, the
central 3 - 5 kpc high density gas core of massive galaxies can survive
from the prompt ram pressure stripping. 
We then perform the similar analysis to the coronae of 10$^{10}$ L$_{\odot}$
galaxies, e.g., those of NGC~3841 and CGCG~97090. The inner stellar potential
profile of CGCG~97090 can be estimated from the \hst\ data. We assume the
following parameters for the initial X-ray corona of CGCG~97090: a corona
radius of 5 kpc, $\beta_{\rm gas}$=0.5, r$_{\rm core, gas}$=0.5 kpc, a
central gas density of 0.01 cm$^{-3}$. These parameters yield a reasonable
X-ray luminosity for the galaxy's L$_{\rm B}$. The assumed X-ray corona is
at a wind or partial wind stage. The critical velocity for
fully stripping is $\sim$ 9.0 $\times 10^{2}$ n$_{\rm ICM}$/6$\times 10^{-4}$
cm$^{-3})^{-1/2}$ km/s. Thus, ram pressure stripping can significantly disturb
or even destroy the very central regions of the diffuse X-ray coronae of
$\lsim$ 10$^{10}$ L$_{\odot}$ galaxies
(e.g., NGC~3841 and CGCG~97090) in this region of A1367.

Continuous stripping by K-H instability and viscosity can be more effective than
the ram pressure stripping (e.g., Nulsen 1982). The mass-loss rate of a galaxy
corona through the K-H instability is (Nulsen 1982):

\begin{eqnarray}
\dot{M}_{\rm KH} &\approx& \pi r^2 \rho_{\rm ICM} {\rm v_{gal}} \nonumber \\
&=& 0.7 (\frac{n_{\rm e, ICM}}{10^{-3} {\rm cm}^{-3}}) (\frac{r}{3 {\rm kpc}})^2
(\frac{{\rm v}_{\rm gal}}{840 {\rm km/s}}) {\rm M_{\odot} / yr}
\end{eqnarray}

The stellar mass loss, in principle, can replenish the gas loss due to the 
stripping processes, if the stellar materials can enter the hot corona phase
efficiently. Using the generally adopted mass loss rate $\dot{M}_*$ = 0.15
M$_{\odot}$ yr $^{-1}10^{10}$ L$_\odot^{-1}$ (Faber \& Gallagher 1976)
and the optical luminosity within the coronae (from the \hst\ data and
GOLD Mine), we estimate the $\dot{M}_{\rm KH} / \dot{M}_{*}$ ratio for
detected coronae: $\sim$ 4.3 at 4.1 kpc radius for NGC~3842, $\sim$ 2.1
at 2.5 kpc radius for NGC~3837, $\sim$ 1.9 at 3.5$''$ radius for NGC~3841 and
$\sim$ 1.8 at 3$''$ radius for CGCG~97090, assuming a velocity of 840 km/s.
The ratios may be still consistent with unity, except for NGC~3842, which
however should have a velocity much less than the assumed 840 km/s. Thus,
the stellar mass loss may help these small coronae survive from the K-H
instability, even without the help of gravity and surface magnetic field to
suppress the K-H instability. Generally to make $\dot{M}_{\rm KH}$ /
$\dot{M}_{*}$ small ($\lsim$ 1), small coronae with high $\dot{M}_{*}$ inside
the boundary are preferred.

The mean free path around these coronae is $\sim$ 10 kpc, if there is no
tangled magnetic field in the ICM to reduce it dramatically. This is larger than
the sizes of coronae so the transport process by viscosity should be saturated.
Nulsen (1982) argued that the saturated thermal evaporation remains dominant
over the transport process by viscosity if both are saturated. As the saturated
thermal evaporation is largely suppressed at the boundary (V01; $\S$3.2), the
saturated transport process by viscosity should also be suppressed, making it
not an important mechanism for the stripping of these coronae.

We conclude that the following factors are favorable to the survival of the
X-ray coronae from gas stripping: well developed galactic cooling core
(associated with most massive galaxies) and small relative velocity to the
surrounding medium (e.g., group / cluster central galaxy).

\subsection{Thermal evaporation \& energy balance inside the coronae}

The survival of cool galaxy coronae in a hot ICM can be used to derive
limits on the efficiency of heat conduction. There are at least two arguments
against the existence of a moving conduction front in the two big coronae.
First, the temperature changes across the boundaries are sharp. Second,
the two big coronae in A1367 are very similar to those in Coma, although
Coma is several Gyr dynamically evolved compared to A1367 and the evaporation
timescales of these coronae are very short ($\lsim$ several 10$^{7}$ yr).
For the measured gas temperature and density across the boundaries, the
mean free path of electron from the hot ICM to the cool ISM is 6 - 9 kpc
for NGC~3842 and NGC~3837, while the mean free path of electrons from the
cool ISM to the hot ICM is 5 - 8 kpc. These values are comparable or
larger than the upper limit on the width of the boundaries ($\sim$ 4 kpc).
Thus, heat conduction at the ISM-ICM boundaries should be saturated. The
saturated conductive heat flux is (Cowie \& McKee 1977):

\begin{eqnarray}
q_{\rm sat} &=& 0.4 (\frac{2kT_{e}}{\pi m_{e}})^{1/2} n_{e}kT_{e} \nonumber \\
&=& 6.8 \times 10^{-4} (\frac{T}{1 {\rm keV}})^{3/2}
(\frac{n_{e}}{10^{-3} {\rm cm^{-3}}}) {\rm \mbox{ }ergs\mbox{ }s^{-1}\mbox{ }cm^{-2}}
\label{eq:cond_sat}
\end{eqnarray}

Because of the large temperature jump across the boundaries, the heat flux
transfered from the hot ICM to the coronae is very large and has to be suppressed,
if the cool coronae are to survive. There is a
temperature gradient inside the coronae (at least for NGC~3842). The transfered
heat flux can be redistributed within the coronae to offset cooling if the heat
conductivity is adequate within the coronae. If the reduced net heat flux
from the hot ICM is balanced by radiative cooling of the coronae, the suppression
factor of heat conduction can be estimated. The net heat flux transfered from the
hot ICM is $\sim$ 2 $\times 10^{43}$ ergs s$^{-1}$ across the 4.1 kpc boundary of
the NGC~3842 corona, and 3 $\times 10^{42}$ ergs s$^{-1}$ across the 2.5 kpc
boundary of the NGC~3837 corona. The radiative cooling flux is only $\sim$ 1.2
and $\sim$ 0.5 $\times 10^{41}$ ergs s$^{-1}$ for NGC~3842 and NGC~3837
respectively. Thus, the heat conductivity has to be suppressed 60 - 170 times at
their ICM-ISM boundaries. Similar estimates can be done for the two smaller
coronae although they are disturbed by the ram pressure. The suppression
factor is 80 - 120 times for them. The heat conduction may be suppressed
by the disjoint magnetic field structure at the boundary.
One possible mechanism to cause the disjoint magnetic field structure
at the boundary is the compression of the galactic magnetic field in the
course of falling into the cluster, which is however a very uncertain
process.

Inside the NGC~3842 and NGC~3837 coronae, since the gas cooling time is shorter
than the Hubble time everywhere (e.g., t$_{\rm cooling}$ = 10 - 40 Myr at their
centers), gas cooling is efficient. The heat conduction within the galaxy
coronae may be able to balance radiative cooling, if heat conductivity is close
to the Spitzer value. V01 tested this idea for the Coma galaxy NGC~4874 and
found that conduction can balance cooling. We test it on the NGC~3842
corona. We first derive the deprojected temperature profile of the NGC~3842
corona. It can be well fitted with T(r)/1 keV = 0.76 + 0.48 (r/10$'')^{1.10}$.
The classical conduction flux can be derived with this temperature profile,
which is consistent with the cooling flux within the 1$\sigma$ uncertainties inside
4 kpc. However, if the conductivity is only 0.1 of the Spitzer value, the
conduction flux is only $\sim$ 0.1 of the cooling flux within the central
2 kpc, and is unable to quench the cooling. The mass flow rates,
$\dot{M} \approx 2\mu$m$_{\rm p}$L$_{\rm X, bol}$/5kT, are 0.51 and 0.34
M$_{\odot}$ yr$^{-1}$ for the NGC~3842 and NGC~3837 coronae respectively.
Stellar mass loss can replenish the X-ray gas. Using the scaling relation
by Faber \& Gallagher (1976), we derive the stellar mass loss rates within
the coronae as $\sim 0.4$ (NGC~3842) and $\sim 0.2$ (NGC~3837) M$_{\odot}$
yr$^{-1}$, which are comparable 
to the mass flow rates. In a summary, the energy balance of the coronae can
be understood by suppressed heat conduction (60 - 170 times) at the boundary
and Spitzer conductivity in the interior. If the interior heat conductivity is
not that high, stellar mass loss can compensate the gas loss from cooling. The
gas inflow from the surrounding ICM should not be important here, as the ICM cooling
time surrounding the NGC~3842 and NGC~3837 coronae is longer than the Hubble time.

\subsection{The effects of rich environments on galaxy coronae}

The properties of galaxy X-ray coronae are significantly affected by the rich
cluster environment. Four other galaxy coronae were found in Coma (V01) and
A1060 (Yamasaki et al. 2002). Similar to those found in A1367, they are small
($\sim$ 3 kpc) and not massive (M$_{\rm gas}$ $\sim$ 10$^{8}$ M$_{\odot}$). The
richness of environment can be represented by the ambient pressure surrounding
the galaxy corona, n$_{\rm e, ICM}$kT$_{\rm ICM}$, which is also proportional
to the average ram pressure that the corona undergoes, since
$\sigma_{\rm cluster} \propto$ T$_{\rm ICM}^{1/2}$. Two easily and robustly measured
properties of galaxy coronae are luminosity and size. In the field and in poor
environments, it is generally found that L$_{\rm X} \propto$ L$_{\rm B}^{\rm n}$
for galaxy coronae, where n ranges from 1.7 to 3 (e.g., Brown \& Bregman 1998;
O'Sullivan et al. 2003). We adopt n=2 and take the normalized rest-frame
0.5 - 2 keV luminosities of the coronae (L$_{\rm X}$/L$_{\rm B}^{2}$) as the
first measure of the coronae. The second measure of the coronae is the
R$_{\rm X}$ / R$_{\rm e}$ ratio, where R$_{\rm X}$ is the size of the X-ray
corona and R$_{\rm e}$ is the effective radius of the galaxy in optical. The
properties of seven galaxy coronae are plotted in Fig. 6.
The host galaxies of these coronae are all luminous galaxies with L$_{\rm B}$
from 0.26 to 1.8 $\times 10^{11}$ L$_{\odot}$. The properties of NGC~1404
are from Machacek et al. (2004). The two faint ones in A1367 are not included because
they are in destruction. Indeed the coronae become smaller (over one order of
magnitude) and less X-ray luminous (over two orders of magnitude) with the
increasing ambient pressure (over almost two orders of magnitude). The trend
is not changed when we change n to 1.7 - 3.

\subsection{Central AGN activity}

The nuclear source can also significantly affect the properties of coronae.
Indeed, there is a radio source ($\sim$ 13 mJy) in NGC~3842 with two radio lobes (Fig.
7)\footnote{Bliton et al. 1998 classified the NGC~3842 radio source as a
narrow-angle tailed (NAT) radio source based on seeming connection between the two
lobes. However, the ``head'' of the source is the eastern lobe rather than
the galactic nucleus, which is against the ``NAT'' argument.}.
The radio lobes are well outside the main body of the gaseous corona.
The symmetrical shape of the X-ray corona implies that any disturbance by
the radio plasma is small. The radio luminosity of the source is $\sim$
10$^{39}$ ergs s$^{-1}$, based on the observed radio properties from NVSS
and FIRST (note the 1.4 GHz flux density on GOLD Mine is about 3 times
larger than those from NVSS and FIRST). The mechanical power of the central
AGN is $\sim$ 10$^{2}$ times larger than the observed radio luminosity based
on the empirical relation derived by B\^{\i}rzan et al. (2004). If the central
source is active with this power for several hundred million years, the
energy released is comparable to the total thermal energy within r$_{\rm cut}$
of the NGC~3842 corona ($\sim 4 \times 10^{56}$ ergs).
We note that there is a much brighter radio source with a similar morphology
in NGC~4874 (Fig. 7), a Coma cluster central galaxy with a 1 keV galaxy corona
(V01). The NGC~4874 radio source is $\sim$ 20 times more
luminous than that in NGC~3842, while its X-ray source is only 70\% as luminous
as that of NGC~3842. Therefore, the central AGN of NGC~4874 has enough energy
to heat its galaxy corona significantly. If the AGN mechanical power, associated
with the observed radio sources, was deposited interior to the coronae, they
will be disrupted or even completely destroyed. Thus,
the co-existence of $\lsim$ 1 keV galaxy coronae with active radio sources implies
that the AGN deposit their energy outside the observed small coronae.
The way that the central AGN release energy outwards may also determine the
existence or size of the X-ray coronae.

\section{Conclusion}

The main conclusions of our study are:

1. This work, in combination with V01 and Yamasaki et al. (2002), demonstrates
that cool X-ray galaxy coronae can survive in rich clusters. However,
they are much smaller (2 - 4 kpc), much X-ray fainter and less massive ($\sim$
10$^{8}$ M$_{\odot}$) than those coronae in poor environments. The X-ray coronae
become smaller and less X-ray luminous with the increasing ambient pressure
(Fig. 6).

2. Heat conduction is suppressed by at least 60 times at the ICM-ISM
interface. Inside the NGC~3842 corona, heat conduction with the Spitzer
value can balance radiative cooling, assuming continous heat flux
from the surroundings with a reduced conductivity.
Stellar mass loss can also compensate the gas loss due to cooling.

3. Generally only the most massive galaxies, with well
developed central cooling cores (e.g., NGC~3842 and NGC~3837), can survive
from gas stripping in the rich environments. The X-ray corona of the central
galaxy, if shielded by the surrounding ICM halo, can also survive in rich
clusters. Galaxies less massive than L$^{\ast}$ (10$^{10}$ L$_{\odot}$)
galaxy, generally with low density coronae in the wind or partial wind
stage, are vulnerable to gas stripping (e.g., NGC~3841 and CGCG~97090).

4. The co-existence of cool X-ray coronae and active radio sources (Fig. 7)
in NGC~3842 and NGC~4874 implies that AGN deposit most of their mechanical power
outside the current galaxy coronae.

\acknowledgments

This research has made use of the GOLD Mine Database, operated by the
Universita' degli Studi di Milano- Bicocca. We acknowledge support from
the NASA contracts NAS8-38248 and NAS8-39073.

\begin{table}
\begin{scriptsize}
\begin{center}
\caption{The physical properties of early-type galaxies with cool galaxy coronae in A1367}
\vspace{0.3cm}
\begin{tabular}{ccccccccc}
\hline \hline
 Name (Type) & V$^{\rm a}$ & L$_{\rm B}^{\rm a}$ & $\sigma_{\rm r}^{\rm a}$ & Counts & T & L$_{\rm X}^{\rm b}$ (soft) & L$_{\rm bol}^{\rm c}$ & L$_{\rm Radio}^{\rm d}$ \\
      & (km/s) & (10$^{10}$ L$_{\odot}$) & (km/s) & (0.5 - 3 keV) &(keV) & (10$^{40}$ ergs s$^{-1}$) & (10$^{40}$ ergs s$^{-1}$) & (10$^{22}$ W Hz$^{-1}$) \\ \hline

NGC~3842 (E) & 6316$\pm$9 & 8.9 & 303 & 1003$\pm$37 & 0.93$^{+0.05}_{-0.06}$ & 6.7 & 12.6 & 1.3$^{\rm e}$ \\
NGC~3837 (E) & 6130$\pm$10 & 2.6 & 315 & 453$\pm$23 & 0.62$^{+0.04}_{-0.03}$ & 2.9 & 5.1 & 0.6$^{\rm f}$\\
NGC~3841 (E)$^{\rm g}$ & 6356$\pm$7 & 1.8 & 209 & 50$\pm$9 & 0.3 - 0.7 & 0.47 & $\sim$ 0.8 & $<$0.1 \\
CGCG~97090 (S0)$^{\rm g}$ & 6148$\pm$179 & 0.84 & 175 & 58$\pm$9 & 0.5 - 1.0 & 0.50 & $\sim$ 0.9 & $<$0.1 \\

\hline\hline
\end{tabular}
\end{center}
\begin{flushleft}
\leftskip 25pt
$^{\rm a}$ From NED and GOLD Mine\\
$^{\rm b}$ Rest-frame 0.5 - 2 keV luminosity of the corona (LMXB excluded for NGC~3842 and NGC~3837)\\
$^{\rm c}$ X-ray bolometric luminosity of the corona (LMXB excluded for NGC~3842 and NGC~3837)\\
$^{\rm d}$ 1.4 GHz luminosity \\
$^{\rm e}$ Based on the results from FIRST and NVSS. The flux listed on GOLD Mine is 3 times
larger. \\
$^{\rm f}$ Based on the results from Gavazzi \& Contursi (1994). However, NGC 3837 was not
detected by NVSS and FIRST, though the listed flux should be easily detected by both of them.\\
$^{\rm g}$ The listed temperatures and luminosities are derived from a single thermal component
fit to the total spectra. The listed luminosities may be over-estimates as the contributions from
LMXB and nuclear source cannot be separated from that of thermal gas.\\
\end{flushleft}
\end{scriptsize}
\end{table}

\begin{table}
\begin{scriptsize}
\begin{center}
\caption{The spectral fits of the NGC~3842 and NGC~3837 coronae and their surroundings$^{\rm a}$}
\vspace{0.2cm}
\begin{tabular}{rcccc}
\hline \hline
Region & Model$^{\rm b}$ & \multicolumn{2}{c}{Parameters} & $\chi^{2}$/d.o.f. \\
       &                 &   T (keV) & Z (solar) & \\ \hline
NGC~3842 (0 - 24.0$''$) & M & 0.98$^{+0.04}_{-0.05}$ & 0.27$^{+0.07}_{-0.06}$ & 52.1/49 \\
 & M+ZB & 0.93$^{+0.05}_{-0.06}$ & 0.34$^{+0.13}_{-0.04}$ & 48.0/48 \\
(0 - 9.6$''$) & M+ZB & 0.92$\pm$0.05 & 0.56$^{+0.27}_{-0.15}$ & 49.2/57 \\
(0 - 1.4$''$) & M & 0.83$^{+0.05}_{-0.04}$ & 0.41$^{+0.38}_{-0.16}$ & 18.7/20 \\
(1.4 - 2.8$''$) & M & 0.86$^{+0.03}_{-0.04}$ & 0.68$^{+0.75}_{-0.26}$ & 20.5/24 \\
(2.8 - 4.7$''$) & M & 1.01$^{+0.06}_{-0.07}$ & 0.68$^{+0.65}_{-0.28}$ & 27.2/26 \\
(4.7 - 9.6$''$) & M+ZB & 1.10$^{+0.18}_{-0.12}$ & 0.15$^{+0.10}_{-0.06}$ & 15.8/23 \\
(9.6 - 24.0$''$)$^{\rm c}$ & M & 1.8$^{+4.2}_{-0.8}$ & 0.0$^{+0.52}_{-0.0}$ & 3.5/6\\
               & M & 2.7$^{+11.5}_{-1.1}$ & 0.32 (fixed) & 4.1/7\\
(24.0 - 76.7$''$) & M & 5.88$^{+0.83}_{-0.59}$ & 0.36 (fixed) & 78.9/79 \\
(76.7 - 112.1$''$) & M & 6.35$^{+0.82}_{-0.72}$ & 0.36 (fixed) & 42.0/70 \\ \hline
NGC~3837 (0 - 17.2$''$) & M & 0.67$^{+0.05}_{-0.03}$ & 0.14$^{+0.05}_{-0.03}$ & 29.2/26 \\
 & M+ZB & 0.62$^{+0.04}_{-0.03}$ & 0.28$^{+0.08}_{-0.06}$ &  19.6/25 \\
(0 - 6.8$''$) & M+ZB & 0.62$^{+0.04}_{-0.03}$ & 0.32$^{+0.12}_{-0.07}$ & 25.7/32 \\
(0 - 2.5$''$) & M+ZB & 0.62$\pm$0.04 & 0.34$^{+0.24}_{-0.08}$ & 15.1/18 \\
(2.5 - 6.8$''$) & M+ZB & 0.65$^{+0.05}_{-0.06}$ & 0.3 (fixed) &  16.6/17 \\
(6.8 - 17.2$''$)$^{\rm c}$ & M & 2.1$^{+3.7}_{-1.0}$ & 0.0$^{+2.0}_{-0.0}$ & 1.7/2 \\
               & M & 2.7$^{+4.6}_{-1.1}$ & 0.32 (fixed) & 2.0/3 \\
(17.2 - 54.1$''$) & M & 4.75$^{+0.83}_{-0.56}$ & 0.36 (fixed) & 52.5/46 \\
(54.1 - 93.4$''$) & M & 5.22$^{+0.61}_{-0.60}$ & 0.36 (fixed) & 60.9/62 \\ \hline
S3 CCD (cluster) & M & 5.29$\pm$0.19 & 0.36$\pm$0.07 & 232.9/174 \\
S2 CCD (cluster) & M & 4.66$^{+0.27}_{-0.18}$ & 0.26$\pm$0.08 & 152.9/122 \\
\hline\hline
\end{tabular}
\end{center}
\begin{flushleft}
\leftskip 25pt
$^{\rm a}$ The profiles are shown in Fig. 4. Within 24.0$''$ of NGC~3842 and 17.2$''$
of NGC~3837, ICM background, determined from regions beyond these radii, is subtracted.\\
$^{\rm b}$ For spectra that a significant hard X-ray excess is present (especially for outer
regions of coronae where LMXB emission is significant), we add a ZBREMSS component with a fixed
temperature of 8 keV. ``M'' represents MEKAL and ``ZB'' represents ZBREMSS.\\
$^{\rm c}$ For these two annuli with poor statistics, Cash statistics is applied rather
than $\chi^{2}$ statistics though the values of $\chi^{2}$ are still cited. \\
\end{flushleft}
\end{scriptsize}
\end{table}

\begin{table}
\begin{scriptsize}
\begin{center}
\caption{The spatial properties of two bright coronae}
\vspace{0.3cm}
\begin{tabular}{cccccc}
\hline \hline
Object & r$_{\rm c}$ & $\beta$ & $r_{\rm cut}$ & n$_{\rm e;0}$ & M$_{\rm gas}$ \\
       & (kpc)       &         & (kpc)         & (cm$^{-3}$)   & 10$^{8}$ M$_{\odot}$ \\ \hline

NGC 3842 & 0.56$^{+0.07}_{-0.08}$ & 0.56$\pm$0.02 & 4.1$\pm$0.3 & 0.26$^{+0.07}_{-0.04}$ & 1.5$\pm$0.6 \\
NGC 3837 & 1.42$^{+0.32}_{-0.20}$ & 1.45$^{+0.46}_{-0.28}$ & 2.5$\pm$0.4 & 0.21$^{+0.06}_{-0.07}$ & 0.59$^{+0.30}_{-0.20}$ \\

\hline\hline
\end{tabular}
\end{center}
\end{scriptsize}
\end{table}

\begin{figure}
\vspace{3cm}
\centerline{{\Huge 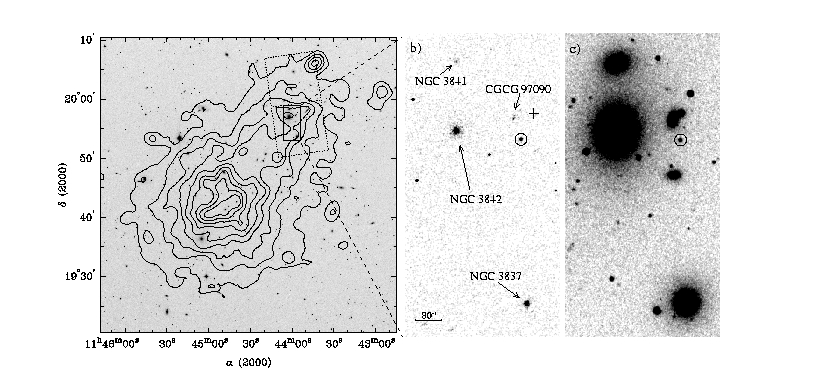}}
\vspace{2cm}
  \caption{a): \rosat\ contours of A1367 superposed on the DSS image. The two
brightest X-ray point sources are excluded, including the brightest one in b)
and c). The two squares (dotted lines) show the S2 (north) and S3 (south) CCD
fields. The small rectangle (solid lines) is the enlarged region on the right.
b): the enlarged region observed by \chandra\ in the 0.5 - 3 keV energy band
(smoothed with one ACIS pixel, in linear scale), with four small galaxy coronae
marked. The cross shows the position of the optical axis, while the circle
marks the brightest X-ray source (a z=0.35 QSO) in the field. c): DSS II image
in the same field as b). The image is in linear scale but the galaxy centers
are saturated to better show the optical sizes of the galaxies.
   \label{fig:img:smo}}
\end{figure}

\begin{figure}
\vspace{-0.7cm}
  \centerline{\includegraphics[height=0.85\linewidth,angle=270]{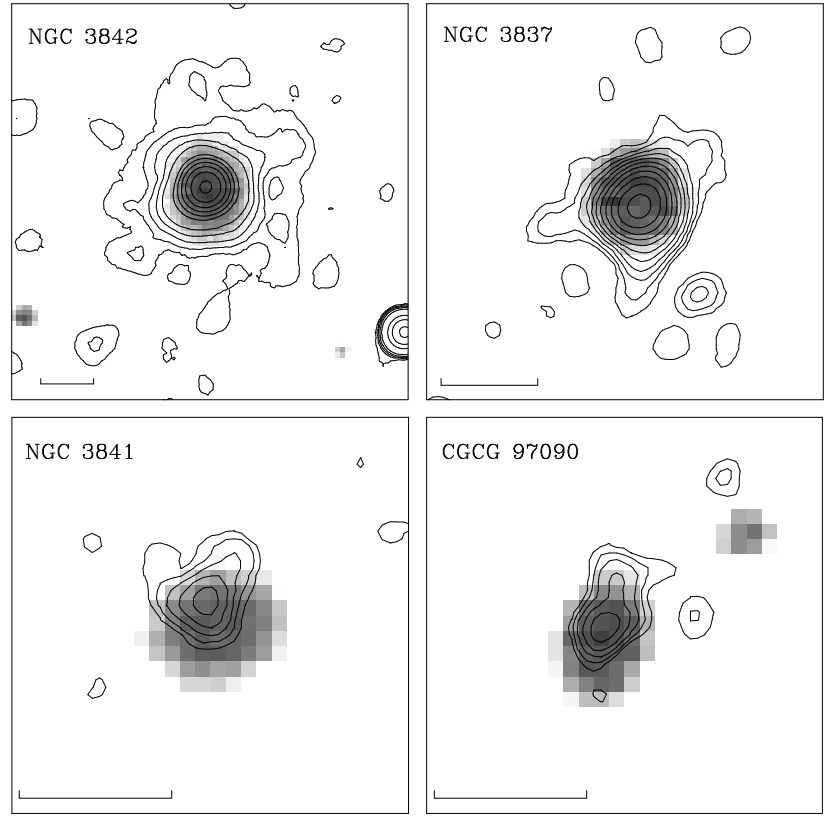}}
\vspace{-0.4cm}
  \caption{X-ray contours of the four galaxy coronae superposed on the DSS images
of their nuclei. The contours levels increase by a factor of $\sqrt{2}$. The peak
surface brightness of the NGC~3841 and CGCG~97090 coronae is 0.9$\times10^{-4}$
cnts s$^{-1}$ arcsec$^{-2}$ and 1.3$\times10^{-4}$ cnts s$^{-1}$ arcsec$^{-2}$
respectively, while the peak surface brightness of the NGC~3842 and NGC~3837
coronae is shown in Fig. 4. The scalebar in each plot represents
10 arcsec. For NGC~3842 and CGCG~97090 which were covered by \hst\ observations,
the X-ray peaks are consistent with the optical peak within 0.1$''$, although
the X-ray emission of CGCG~97090 is extended to the NW. The X-ray peak
of NGC~3837 is $\sim 1.1''$ south of the optical peak, which is still consistent
with the positional uncertainty of DSS and \chandra. The X-ray peak of NGC~3841
is displaced to the north of the optical peak ($\sim 1.9''$).
   \label{fig:img:smo}}
\end{figure}

\begin{figure}
\vspace{-0.5cm}
  \centerline{\includegraphics[height=0.7\linewidth,angle=270]{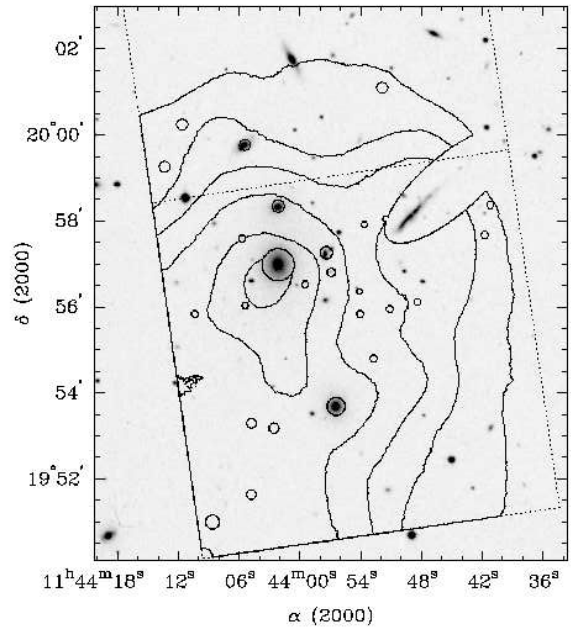}}
  \caption{Diffuse cluster emission (contours, linearly spaced) superposed
on the DSS image. The small coronae and point sources are masked (the
circles and the large ellipse) before smoothing (35$'' \sigma$).
The dotted lines delineate the S3 CCD and part of the S2 CCD (see
Fig. 1 to compare with the PSPC image). There is a weak ICM enhancement
around NGC~3842 and the cluster emission is elongated to the SE.
   \label{fig:img:smo}}
\end{figure}

\begin{figure}
\vspace{-0.2cm}
  \centerline{\includegraphics[height=1.0\linewidth,angle=270]{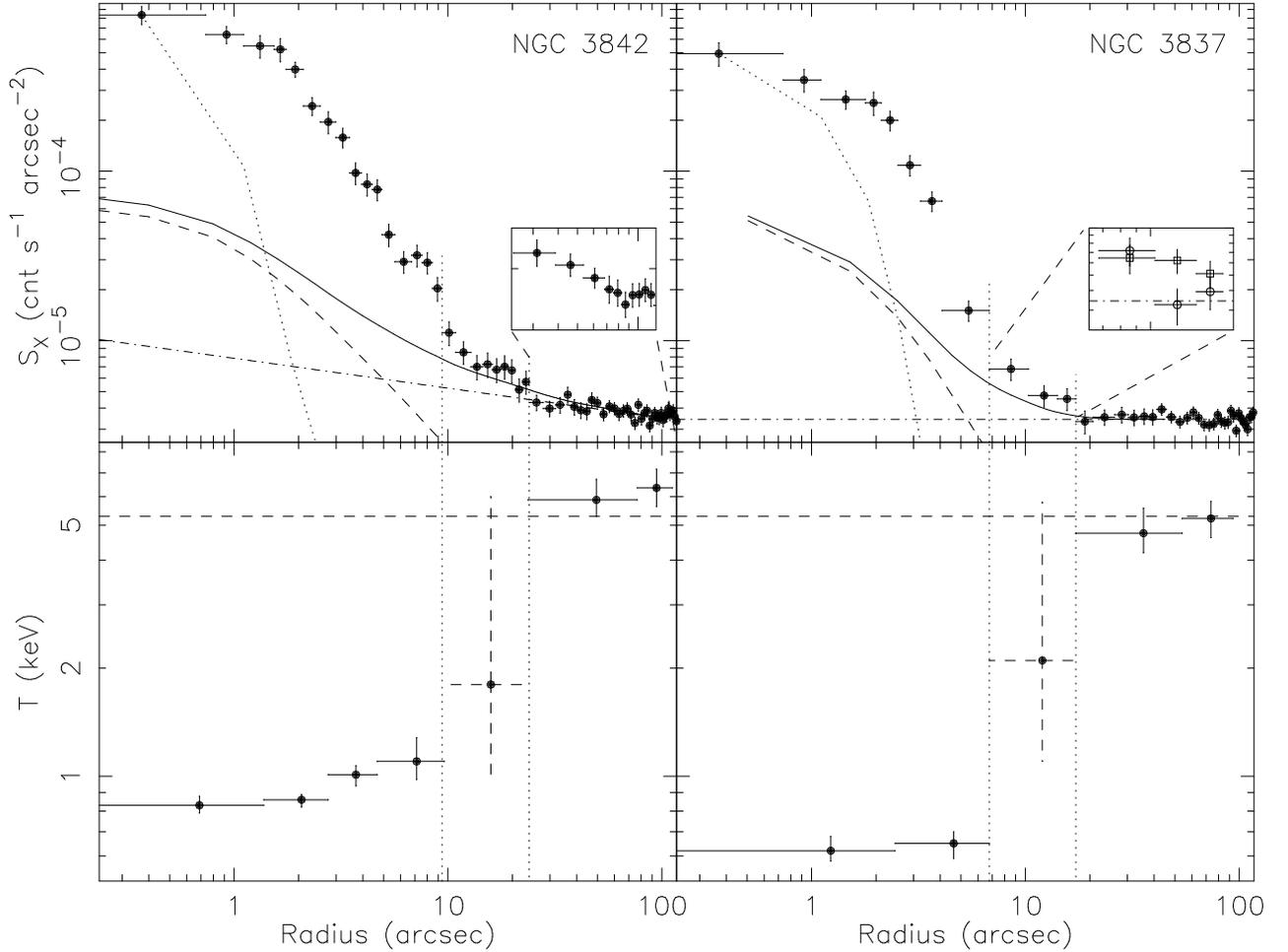}}
\vspace{-0.4cm}
  \caption{The 0.5 - 2 keV surface brightness and temperature profiles of the
NGC~3842 ({\bf Left}) and the NGC~3837 coronae ({\bf Right}). In the surface
brightness plot, the dotted line represents the local \chandra\ PSF. The
dashed-dotted line is the local background, which is slightly peaked around
NGC~3842 (see the small rectangular insert) but constant around NGC~3837.
The dashed line is the expected LMXB X-ray light profile
if it exactly follows the optical light. The LMXB X-ray
light profile is normalized to match the predicted total LMXB X-ray luminosity
(see text). The solid line is the combination of the LMXB component and the
local background. The regions between the vertical dotted lines are where the
LMXB component may be important (see $\S$2.3). As shown in the small rectangular
insert, the NGC~3837 X-ray emission to the south (open squares) is more extended
than that to the north (open circles), which may be caused by stripping of the
outer envelope. In both temperature profiles, the gas temperature increases
abruptly from the galaxy virialized temperature to the ICM temperature (5 - 6
keV) across the boundary. The horizontal dashed line represents the best-fit ICM
temperature of the whole S3 CCD field. The measured temperatures in the ``plateau''
regions (between the dotted lines) are still consistent with the value expected
for LMXB, but some contribution from cool thermal gas is possible (see text).
   \label{fig:img:smo}}
\end{figure}

\begin{figure}
  \centerline{\includegraphics[height=0.38\linewidth]{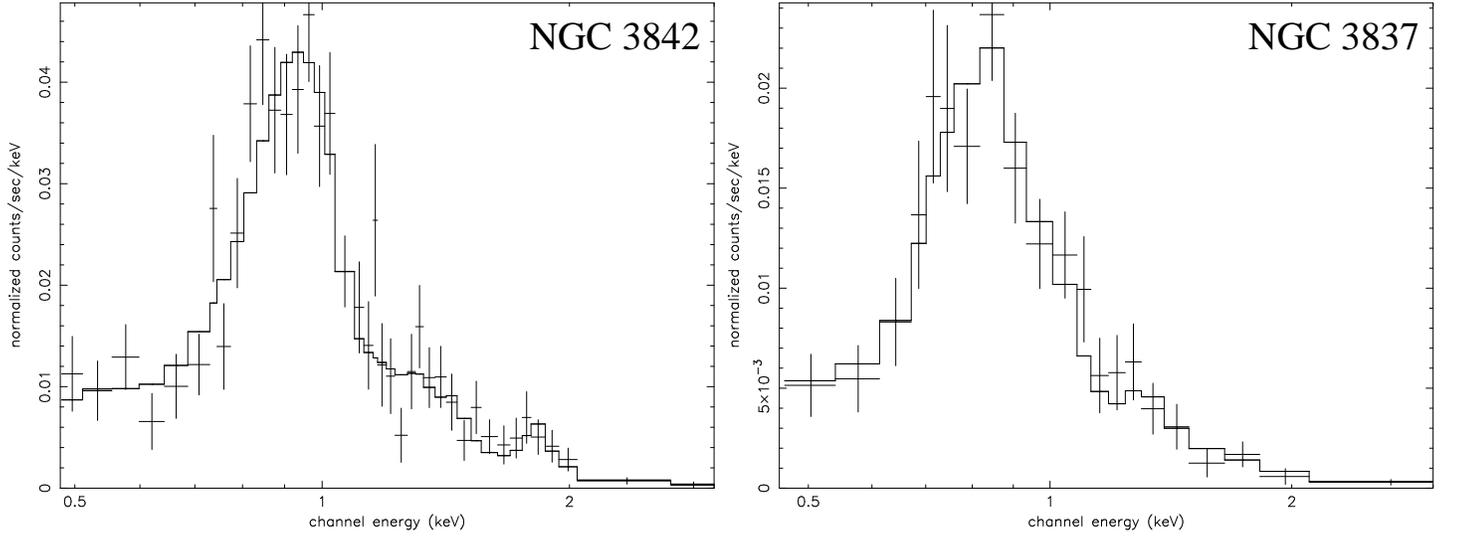}}
  \caption{The global spectra of the NGC~3842 and NGC~3837 coronae with the
best-fit ``VMEKAL + ZBREMM'' model. The blend
of iron L lines is significant in both spectra. The blend centroid of NGC~3842
is at higher energy than that of NGC~3837, which implies a higher temperature
of the NGC~3842 corona than the NGC~3837 corona. Si He$\alpha$ line can also
be seen around 1.8 keV in the spectrum of NGC~3842.
   \label{fig:img:smo}}
\end{figure}
\newpage

\begin{figure}
\vspace{-4cm}
  \centerline{\includegraphics[height=0.95\linewidth,angle=270]{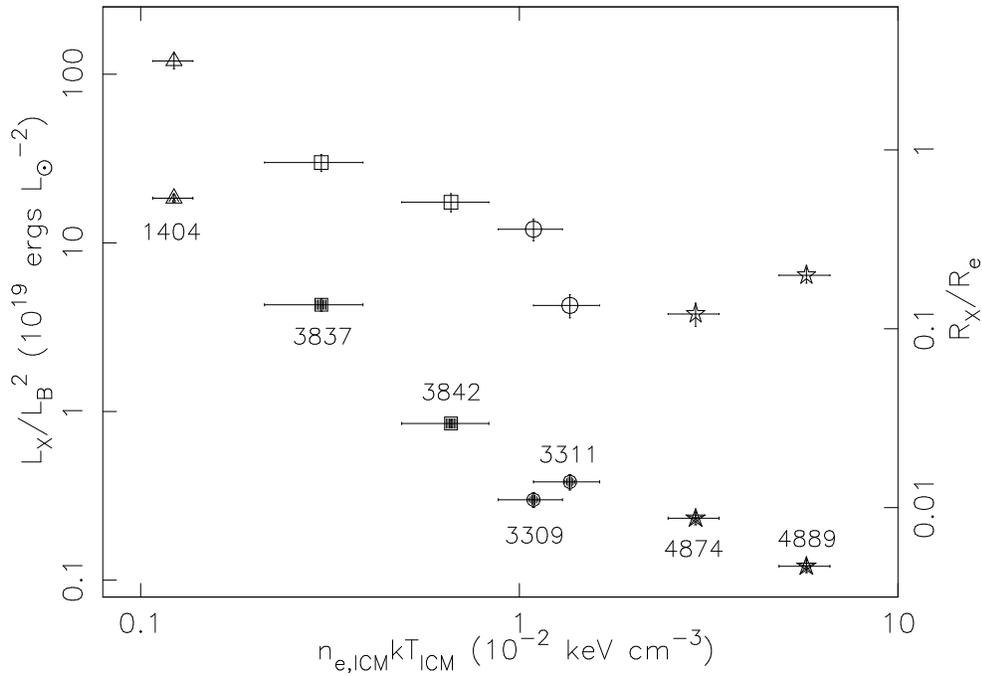}}
  \caption{The effects of environment (represented by n$_{\rm e, ICM}$
kT$_{\rm ICM}$, the ambient pressure) on the properties of the
X-ray coronae in Coma (stars), A1060 (circles) and A1367 (squares). One galaxy
in a poorer environment (NGC~1404 in Fornax cluster) is also included. The solid points
represent the 0.5 - 2 keV luminosities normalized by L$_{\rm B}^{2}$, while the
open symbols represent the sizes of the X-ray coronae normalized by the effective
radius (R$_{\rm e}$) of the galaxy. Both properties generally decrease with the
ambient pressure, reflecting the effects of the environment. This plot is only
for the most luminous early-type galaxies where galactic cooling core can be
developed, while low mass galaxies cannot hold their coronae in high pressure
environment.
   \label{fig:img:smo}}
\end{figure}

\begin{figure}
\vspace{-7cm}
  \centerline{\includegraphics[height=1.15\linewidth,angle=270]{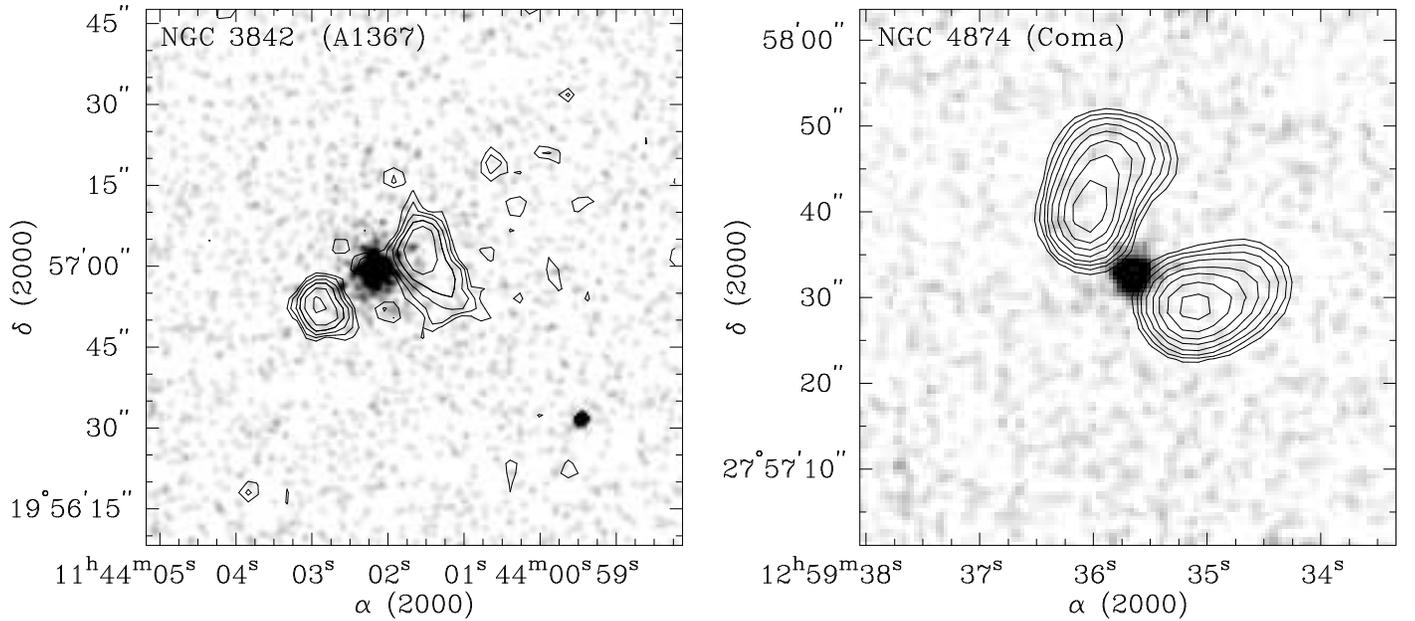}}
\vspace{-0.5cm}
  \caption{{\bf Left}: the 20 cm VLA FIRST contours of NGC~3842 superposed
on the \chandra\ image (smoothed by one ACIS pixel). A similar VLA map
obtained by Owen \& Ledlow (1997) shows that both radio lobes originate
from the nucleus. However, the bulk of the radio emission appears outside the
main body of the X-ray gas. The existence of an active radio source and
a cool X-ray corona in NGC~3842 imply that the AGNs release energy mostly
at large radii, outside of the existing coronal gas. The relatively
symmetrical radio morphology implies no significant ram pressure on the
corona. {\bf Right}: a similar source in the Coma cluster (NGC~4874; the 20
cm VLA FIRST contours superposed on the \chandra\ image). It also has a 1 keV
X-ray corona (V01) and the radio source is $\sim$ 20 times more luminous than
that in NGC~3842. The radio morphology may suggest some motion of the galaxy
relative to the surrounding medium.
   \label{fig:img:smo}}
\end{figure}

\end{document}